\documentstyle[aps,manuscript]{revtex}

\begin{document}               

\title{
{\normalsize {\em LSU Space Science/Particle Astrophysics Preprint -- 11/30/99}}\\
\mbox{  }\\ Charged particle production in the Pb+Pb system at \\
158 GeV/c per nucleon}

\author{P. Deines-Jones,$^{1*}$ M. L. Cherry,$^1$ A. Dabrowska,$^2$ R. Holynski$^2$, D. 
Kudzia$^2$, B. S. Nilsen$^{1x}$, A. Olszewski$^2$, M. Szarska$^2$, A. Trzupek$^2$, C. J. 
Waddington$^3$, J. P. Wefel$^1$, B. Wilczynska$^2$, H. Wilczynski$^2$, W. Wolter$^2$, B. 
Wosiek$^2$, and K. Wozniak$^2$\\
\mbox{  }\\
{\it 1  Louisiana State University, Baton Rouge, Louisiana\\
2	Institute of Nuclear Physics, Krakow, Poland\\
3	University of  Minnesota, Minneapolis, Minnesota\\ 
$^*$ Present address: NASA Goddard Space Flight Center, Greenbelt, Maryland\\
$^x$ Present address: Ohio State University, Columbus, Ohio \\}}

\maketitle

\begin{abstract}              
Charged particle multiplicities from high multiplicity interactions of 158 GeV/nucleon Pb 
ions with Pb target nuclei have been measured using nuclear emulsion chambers.  
The characteristics of these interactions have been compared to those of central interactions 
of 200 GeV/nucleon proton, O, and S beams on silver or bromine targets and those from 
simulations using the FRITIOF 7.02 and VENUS 4.12 Monte Carlo event generators.  Multiplicities 
of Pb+Pb interactions in the central region are significantly lower than predicted by either 
model.  We examine the shape of the pseudorapidity distribution and its dependence on 
centrality in detail, because in this symmetric system the participant projectile target masses 
are independent of centrality, so any dependence of the shape on centrality must therefore be a 
dynamical effect.  VENUS, the only one of the two models which attempts to incorporate reinteraction 
phenomena, predicts a narrowing of the pseudorapidity distributions for the highest multiplicity 
events, which we do not observe. In general, FRITIOF produces better fits to the data than does 
VENUS.\\
\mbox{  } 
\end{abstract}

\section{ INTRODUCTION}\label{intro}

	The superposition model of nucleus-nucleus (AA) interactions has been highly successful 
in describing the general features of particle production in high energy heavy ion systems.  With 
the availability of beams of $^{208}$Pb at the CERN SPS, superposition can now be tested over 
two orders of magnitude in projectile or target mass from pp to Pb+Pb, and three orders of 
magnitude in the number of nucleon-nucleon (NN) collisions.  The Pb+Pb system provides 
nearly the largest reaction volume achievable, and the highest energy densities attainable until 
RHIC and LHC begin colliding heavy ion beams.  In this paper, we report Pb+Pb results from 
the Krakow-Louisiana-Minnesota (KLM) emulsion chamber exposures (EMU-13) at the SPS in 
December 1994.  Previous results from EMU-13, and a detailed description of the experiment, 
have been presented in refs. [1,2].

	Currently, a major emphasis in this field is the search for non-superposition processes in 
rare events or at high energy densities.  In order for a state such as a quark-gluon plasma to be 
produced, superposition must break down via some thermalization process, such as reinteraction.  
To detect and understand events in which plasma or similar collective behavior occurs, it may 
well be necessary to have a quantitative understanding of `ordinary' superposition and 
reinteraction physics, especially if events exhibiting collective behavior are  rare.  Despite the 
fact that the SPS operates in an energy regime of high nuclear transparency, the Pb+Pb system is 
large enough that one might reasonably expect reinteraction to have an observable effect on the 
distribution of produced particles in the most central events.

	At the moment, we lack highly precise predictors of particle production and angular 
distributions in AA interactions.  The simplest and historically the first such predictor is the 
wounded nucleon model [3], which assumes that the multiplicities $n_{AA}$ scale with the 
average number of participating or `wounded' nucleons $W$ and with the average proton-proton 
multiplicity $n_{pp}$  at an equivalent energy per nucleon:  $n_{AA}(E) = \frac{1}{2} Wn_{pp}(E)$.  
This contrasts with the naive expectation that the multiplicities should scale with the number of 
NN interactions, which would result in multiplicities much larger than those from the wounded 
nucleon calculation, since a participating projectile nucleon typically interacts with several target 
nucleons ($\sim 5$ in central Pb+Pb interactions).  At SPS energies, the wounded nucleon model 
predicts multiplicities which are systematically lower than those observed [4-6], but only by $\sim 
15\%$ for all measured systems from pp to Pb+Pb.  In fact, the current generation of Monte Carlo 
codes are no better at predicting multiplicities in ultra-heavy systems than the wounded nucleon 
model, although they do of course provide much more comprehensive predictions of angular 
distributions, transverse momenta, secondary particle species, etc.

	At high energy, individual projectile nucleons should typically interact with several target 
nucleons before re-hadronizing outside the target nucleus.  Thus, the physical implication of the 
relative success of the wounded nucleon model is that interactions of hadronic excitations 
(`reinteractions' of the collision products emerging from the initial NN interactions) appear to 
contribute little to the final-state multiplicity.  To look for dynamical effects besides this 
interesting but well-known `wounded nucleon effect', and to study reinteraction effects in detail, 
one needs to characterize the multiplicities of AA systems carefully.  This paper characterizes 
the measured Pb+Pb multiplicities and angular distributions and compares them to lighter 
systems and to predictions from FRITIOF 7.02 [7] and VENUS 4.12 [8].  Other experiments, 
both electronic [9-13] and emulsion [14-16], have reported measurements of high energy Pb+Pb 
interactions, including results on multiplicities, strangeness and $J/\psi$ production, flow, 
intermittencies, etc.  Measurements made with nuclear emulsions have the unique advantage that 
their excellent position resolution makes it possible to study the extreme forward direction where 
the projectile spectators appear [1].  In this study we separate the central region, where effects 
unique to AA interactions are thought to occur, from the spectator region.  Both regions are 
examined for predicted signs of reinteraction.

	In Sec. II we calculate the number of wounded nucleons $W$.  Then in Sec. III we describe 
the experiment and analysis procedure, and use the measurement of forward charge to 
demonstrate the validity of the calculation of $W$.    At the same time, we derive a value for the 
charged particle multiplicity $n_0$ in the absence of spectators, a value which appears to be 
lower than expected from the simulations.  In the framework of the wounded nucleon model, 
$n_0$  is proportional to $W$.    The multiplicity per wounded nucleon $m$ is therefore 
independent of $W$ and as a result also independent of impact parameter, so that $m$ is an 
appropriate quantity to test for effects unique to AA collisions.  In Sec. IV, we compare the 
measured value of m for central Pb+Pb collisions to the FRITIOF and VENUS simulation 
results, to the results of pN interactions at similar energies, and to the results for other AA 
systems.  In Sec. V we look at the detailed shapes of the pseudorapidity distributions and find a 
distribution which is broader than predicted by VENUS (either with or without reinteraction) but 
similar to that predicted by FRITIOF without reinteraction.  We then look in the forward region 
and use the measured shape of the produced particle distribution to separate out the spectators 
and derive an average transverse momentum for the spectator protons.

\section{CALCULATION OF THE NUMBER OF PARTICIPANT NUCLEONS}\label{II}

	In comparing multiplicities in systems of different sizes,  a convenient quantity is the 
multiplicity per participating nucleon $m=n/W$.  In the context of the superposition model, this 
expression factorizes the observed multiplicity $n$ into two quantities $W$ and $m$.  $W$, the 
number of wounded nucleons, contains all the geometrical effects, i.e., the effects of nuclear 
radii, density, and impact parameter.  The other  factor, $m$, depends only on interaction 
dynamics.  The calculation of $W$ from the experimental data is described here.

	For inclusive datasets, the number of participants, or wounded nucleons, is given in terms 
of interaction cross sections $\sigma$   by [3]

\begin{equation}
W = A_T \frac{\sigma_{NP}}{\sigma_{PT}} + A_P \frac{\sigma_{NT}}{\sigma_{PT}}
\end{equation}

\noindent Here the subscript $P$ means projectile, $T$ means target, and $N$ denotes an 
individual nucleon, so that $\sigma_{PT}$ is the total inelastic hadronic cross section for the 
projectile nucleus interacting with the target [17 - 19], and $\sigma_{NP}$ and $\sigma_{NT}$ 
are the corresponding nucleon-nucleus cross sections.  The first term in (1) is the number of 
wounded target nucleons $W_T$ and the second the number of wounded projectile nucleons 
$W_P$.  To compute $W$ for central data samples, one uses cross sections which are functions 
of the maximum impact parameter $b_{max}$ of the sample. The cross sections are computed 
with a Glauber calculation [20], using the inelastic hadronic cross sections and the nuclear 
density functions of the target and projectile.  The maximum impact parameter of the data 
sample is derived from the partial cross section for producing events in the sample,

\begin{equation}
\sigma_{part} = \pi b_{max}^2 = \sigma_{PT} \frac{N_{cent}}{N_{tot}}
\end{equation}

\noindent where $N_{cent}$ is the number of central events in the data sample and $N_{tot}$ is 
the total number of hadronic AA interactions. $N_{tot}$ is derived either from a minimum bias 
scan of the emulsions or is calculated from the beam count and the total AA cross section.  The 
Glauber calculation gives the number of participant or `wounded' projectile and target nucleons 
$W_P$ and $W_T$.  One can also calculate the number of target and projectile interactions 
$\nu_T$ and $\nu_P$ with the same formalism by assuming that the cross section for nucleons 
which have been excited by a previous interaction is the same as for unexcited nucleons, 
$\sigma_{NN}$.  This assumption is not necessary in the calculation of the number of 
participants $W=W_T+W_P$, which depends (through $\sigma_{NP}$, $\sigma_{NT}$, and 
$\sigma_{PT}$) only on $\sigma_{NN}$ and the nuclear density functions. In this analysis we 
use   numbers of participants and numbers of collisions derived from the FRITIOF simulations 
[7] of the nuclear collision geometry for the specified maximum impact parameters (Eq.2).  The 
sample of central events discussed here corresponds to impact parameters less than 5 fm, 
compared to a maximum of $2.4(208)^{1/3} \sim 14$ fm for Pb+Pb.  At $b$ =  5 fm, the values 
of $W$ derived from different density functions differ by $\sim 3\%$ or less [21].  The deviation 
increases for larger impact parameters.

\section{EXPERIMENTAL AND ANALYSIS PROCEDURE}\label{III}

	The data chosen for this analysis consist of central interactions on targets at least as heavy as 
the projectile.  In these systems, the multiply charged spectator fragments that remain after the 
interaction provide an indication of the centrality of the collision.  The measurements are taken 
from two different kinds of emulsion experiments.  Data from 200 GeV p+Ag/Br [22], O+Ag/Br 
[23], and S+Ag/Br [23,24] collisions were measured in emulsion stacks, in which the detector 
was the target, and which consequently had 4$\pi$ angular acceptance.  A sample of 170 events 
from 158 GeV/nucleon Pb+Pb collisions obtained in CERN experiment EMU-13 was measured 
in chambers with Pb targets and thin emulsion plates exposed perpendicular to the beam[1].  The 
Pb chambers were designed specifically so that charge could be measured in a small inclined 
stack at the downstream end of the chambers [1,2].  In the chambers we measure only the 
particles in the forward cone, $\theta \leq 0.11$ radian, corresponding to a pseudorapidity $\eta 
\geq 2.9$.  This cone includes the peak of the pseudorapidity distribution.  Chambers allow the 
use of targets other than emulsion, and also present less material to induce secondary interactions.  
In both emulsion chambers and stacks, individual particles can be measured even in the extreme 
forward region, and the sub-micron resolution makes it possible to  identify the individual tracks 
of these particles and to measure their individual charges.

	In the stacks, heavily ionizing particles (those with ionization greater than 1.4 times 
minimum) are distinguished from relativistic shower particles, and the numbers of each, $N_h$ 
and $n_s$, are recorded.  Two selections on the minimum bias sets are made to select central 
events on heavy targets [5,24]:  interactions on the Ag or Br nuclei in emulsion are selected by 
choosing those events with $N_h > 15$ heavy tracks produced by slow particles from the target, 
and central events are chosen by selecting those events with no multiply charged beam fragments 
($n_{fr}=0$).  

	In the analysis of the Pb chambers, no minimum bias scan was performed.  Instead, a 
sample of high multiplicity events was selected by visually scanning for large events.  This scan 
efficiently detected events with observed charged multiplicities $> 600$ (corresponding to total 
charged multiplicities larger than about 1000) together with an incomplete sample of events with 
smaller multiplicities [1].  From the measured number of incident primaries and the assumed 
total hadronic Pb+Pb cross section, $\sigma_{Pb+Pb} = 6.9 \pm 0.5$ barns, we estimate that we 
have measured $(23 \pm 4)\%$ of the hadronic interactions.  In 170 events we have fully 
measured all the particles in the pseudorapidity cone $\eta \geq 2.9$ using an automated 
measurement system developed at LSU [25,26].  The measurements of the multiplicities of these 
events have a systematic uncertainty of 3$\%$.  

       The charge modules included in the EMU-13 chambers enabled us to make a  measurement 
of the individual charges of the forward emitted projectile fragments [2]. These measurements 
were made in the most downstream module of the chamber by counting $\delta$-rays along the 
particle tracks. The $\delta$--ray counts were taken on a track length of about 3 mm, giving a 
charge uncertainty of the order of 1-2 charge units for fragments with charge Z $< 30$. These 
charge measurements were completed for 152 of the central Pb+Pb collisions. In the remaining 
18 events, the produced fragments missed the charge section of the chamber, making individual 
$\delta$--ray counts impossible. For these events we assigned charges to multiply charged 
fragments by using the darkness measurements which were made for the all recorded particles 
downstream from the interaction vertex. These darkness measurements were used in our 
preliminary analysis of central Pb+Pb collisions [1] to separate singly charged particles from the 
multiply charged projectile fragments, and correlate nicely with the charge estimates obtained 
from the $\delta$--ray counts, as is shown in Fig.~\ref{fig1}. Therefore we used the parameterization of 
the track darkness vs. charge shown in Fig.~\ref{fig1} to assign charges to fragments produced in the 18 
events which did not reach the charge section of the chamber. This parameterization was also 
used to estimate charges for 7 fragments recorded in our sample of 170 events, which underwent 
a secondary interaction in the upstream elements of the chamber. Over all the analyzed events, 
we observed 540 helium fragments (Z=2) and 175 heavier fragments. The average charge of 
those fragments with Z $> 2$ is small, $5.3 \pm 0.3$, as compared to $16.1 \pm 0.6$ measured in 
inclusive Pb+Pb interactions [2]. The frequency of projectile fragments with Z $> 2$ is also lower 
in the selected central collisions, about 1 fragment per event, while in inclusive interactions we 
observe on average twice as many such fragments per collision. The measurements of the 
individual charges of the projectile fragments allow us to define for each analyzed event the 
quantity $Z_b = \sum_{i=1}^{n_{fr}} Z_i$, the total charge bound in multiply charged fragments. 
This quantity is proportional to the size of the projectile spectator remnant, and therefore is a 
good measure of the impact parameter of the collision. Details of the chamber measurements and 
the automatic measurement system are presented elsewhere [1,2,22-26].

	The measured multiplicities of the Pb+Pb events must be scaled by an angular acceptance 
factor to calculate the total produced multiplicities $n_{prod}$. To determine the acceptance 
factor, pseudorapidity space is divided into three intervals: $\eta < 2.9$, in which multiplicities 
are not measured and must be estimated; $2.9 \leq \eta < 6$, in which all events are completely 
measured, and which is virtually spectator-free; and $\eta \geq 6$, which is also measured but 
which contains spectators as well as produced particles.  In the interval $2.9 \leq \eta < 6$,   the 
shapes of the pseudorapidity distributions are independent of the events' overall multiplicity and 
are reproduced quantitatively by FRITIOF.  (See ref. [1] and Section 4.)   We therefore use a 
single acceptance multiplier derived from FRITIOF for events of all multiplicities: 
$n_{prod}/n_{2.9-6} = 1.80$.  We note that we obtain the same multiplier if we use VENUS 
instead of FRITIOF, even though VENUS produces narrower distributions. This is a 
consequence of the symmetry of the rapidity distribution for the 
Pb+Pb system.  Because we measure pseudorapidity $\eta$ rather than rapidity $y$, there is an 
uncertainty in the position of the pseudorapidity peak, due to the non-zero mass and transverse 
momentum of the produced particles.  This shift $\eta_{peak}-y_{peak}$ is observed to be $0.3 
\pm 0.1$ units of pseudorapidity.  The 0.1 uncertainty in the pseudorapidity corresponds to an 
uncertainty in the multiplier of $2.5\%$.  The overall uncertainty in the Pb+Pb estimates of 
multiplicity, including the $3\%$ measurement uncertainty, is $4\%$.

	The Pb+Pb data samples used in the analysis are shown in Table I.  We tabulate the mean 
bound charge $<Z_b>$, the mean singly charged and fragment multiplicities $<n_{prod}>$ and 
$<n_{frag}>$, and the forward $(\eta \geq 6)$ charge $<Q_{\eta \geq 6}>$ for a set of samples 
selected by different values of $Z_b$.  Note that these event samples are  independent.  The 
sample of events with $n_{prod} \geq 1000$ (corresponding to $9.7\%$ of the total interactions) 
is compared to data from interactions of lighter 200 GeV/nucleon projectiles on silver or bromine 
(Ag/Br), summarized in Table II, where the average number of intranuclear interactions in the 
events covers the range from 4.2 for p+AgBr to 767 for Pb+Pb. 

	In the samples of central and semi-central events in Table I, there are only small numbers 
of spectator protons in the projectile region [1].  Almost all the shower particles in this region are 
`produced' in the sense that they are either created or are participant protons, which may 
originate from either incident protons or neutrons.  In the stack data sets, heavily ionizing 
particles are found in the large-angle target region and are excluded from the analysis in order to 
remove fragments and spectator protons coming from the target spectator.  This cut also removes 
a few produced particles, estimated from VENUS and FRITIOF to be of the order of $2-3\%$ of 
the charged particle multiplicity.  Since the Pb+Pb analysis is restricted to the region $\eta \geq 
2.9$, heavy target fragments are almost completely excluded from these Pb+Pb data.

	In both the stacks and the chambers, the centrality selection is made by cutting on the 
number and charge of beam fragments, which determine $Z_b$.  In the case of the Pb + Pb 
chamber measurements, however, because the beam and target are of equal size, $Z_b$ is a 
variable centrality selection which is more sensitive to the centrality than in the other, 
asymmetric, systems.  Fig.~\ref{fig2} shows the multiplicities and the dispersions of the multiplicity 
distributions as a function of $Z_b$. In the simulations, the $Z_b$ selection is modeled by 
choosing a range of impact parameters corresponding to the number of events in the sample (Eq. 
2).  The FRITIOF and VENUS points are therefore plotted for fixed values of impact parameter 
$b$.  The data points are plotted for the corresponding values of $Z_b$.  For both the data and the 
simulations, the event multiplicities  decrease with increasing $Z_b$, demonstrating that events 
with smaller charge bound in the projectile fragments do indeed have a larger number of 
participants, and on average are more central.  The FRITIOF and VENUS events follow the 
same trend as the measured events, but with higher multiplicities overall.  

	The difference in multiplicity in Fig. 2(a) between the measurements and the simulations is 
at least partly due to the different selection criteria applied to the two data sets.  Fig. 2(b) shows 
the standard deviations $\sigma$ of the multiplicity distributions in each data sample.  The 
measured distributions are broader than those predicted by the simulations, at least partly due to 
our modeling assumption that samples with a particular charge bound in projectile fragments 
correspond to a well-defined range of impact parameters.  This assumption is to some degree 
physically unrealistic.  Instead, we expect there to be some overlap in impact parameters between 
samples of different $Z_b$, and consequently the measured data points correspond to broader 
impact parameter ranges than do the corresponding simulated events. This deviation from the 
ideal selection assumed for modeling purposes has an impact on the results in Fig.~\ref{fig2}.  
Nevertheless, Fig.~\ref{fig2} clearly shows that $Z_b$ correlates reasonably well with impact parameter.

	The symmetry of the Pb+Pb system, combined with the high multiplicities of these events, 
allows us to test our Glauber calculations and at the same time look in some more detail at the 
comparison of measured and calculated multiplicities.  In Fig.~\ref{fig3}, we plot the total charge 
contained in the cone $\eta \geq 6$, which includes essentially all of the spectators, against the 
multiplicity $n_{prod} = 1.80 n_{2.9-6}$ for all the analyzed events. Superimposed on the data 
above $n_{prod} \geq 1000$ (where the selection is close to $100\%$ efficient) are the averages 
for samples with different values of $Z_b$ (solid circles), i.e., centrality.  The charge intercept of 
a fit to the $Z_b$ points (upper line) is $79 \pm 3$,  consistent with the charge of the Pb beam, 
demonstrating that indeed all the spectator charge is contained in the $\eta \geq 6$ cone, as 
asserted above. For each sample, the calculated spectator charge $82(A-W_P)/A$ (where $A$ = 
208 is the projectile atomic number) has been subtracted from the measured total charge to 
obtain the average produced multiplicities in the $\eta \geq 6$ cone (large triangles).  This charge 
grows with the multiplicity in a manner consistent with direct proportionality (cf. Section 4).  
This scaling is observed in the Pb+Pb system at lower, spectator-free pseudorapidities than in the 
p, O, and S+AgBr systems, and is another consequence of the Pb+Pb system's symmetry.  This 
simply means that the shape of this system's pseudorapidity distributions are independent of 
multiplicity (and centrality).  The forward charge data are consistent with the assumption that 
this behavior also holds in the spectator region.  It also suggests that event multiplicity is directly 
proportional to the number of participants, as expected. 

	The intersection of the two fitted lines at $n_{prod} = 1584 \pm 60$ determines the 
multiplicity\footnote{Head-on Pb+Pb events have on average about 9 spectator protons, and therefore 
have slightly lower multiplicities than zero-spectator events.}
  of events with no spectator protons $n_0$.  This value agrees with the value 
determined directly from the $n_{prod} \geq 1000$  sample, $2A/W <n_{prod}> = 1545 \pm 53$. 
Thus, the forward charge measurements, the produced particle measurements, and the calculated 
number of participants together form a tightly consistent picture.  If our calculations of the 
numbers of participants are systematically wrong by as little as $\sim 4\%$, our two methods of 
determining $n_0$ will no longer agree.  Fig.~\ref{fig3} therefore indicates that our participant 
calculations are correct within the stated errors.\footnote{This conclusion depends on the 
assumption that produced multiplicity can really be plotted as a straight 
line through the origin as in Fig.~\ref{fig3}.  Although plausible, this assumption should be tested by
extending the data points in Fig.~\ref{fig3} to the left (i.e., to more peripheral events). Our scanning 
procedure, which was not fully efficient for detecting events with multiplicities smaller than 1000, 
makes this test impossible. Nevertheless, the good fit of the triangles in Fig.~\ref{fig3}
 with a straight line 
through the origin strongly suggests that this assumption is reasonable.  The linearity assumption (as 
suggested by the wounded nucleon model) is discussed and justified further in sec. V.} 

	The spectator-free multiplicity measured here $(1584 \pm 60)$ agrees well with the value 
of $1550\pm120$ determined from the preliminary analysis in [1], and is lower than the value of 
1850 expected from the Monte Carlo simulations.  We note that results reported by Stenlund et 
al. [14] have also suggested a lower central Pb-Pb multiplicity than expected. We have repeated 
our analysis by fitting to points chosen on the basis of the number of alphas rather than $Z_b$, the 
number of shower particles $n_s$, and with different radial density distributions, and obtain 
essentially identical results in all cases.  The conclusions that the analysis is self-consistent and 
that the spectator-free multiplicity is lower than expected appear to be robust and independent of 
the detailed analysis.   

	The advantages of emulsion chambers over stacks are that they allow us to choose our 
target and minimize secondary interactions.  The main shortcomings of chambers, namely their 
limited angular acceptance and the relative difficulty of performing minimum bias scans, turn out 
to be only minor disadvantages for the study of central events because of the symmetry of the 
Pb+Pb system and the good statistics in individual events. Symmetric projectile-target systems, 
such as the Pb+Pb system, are unique in providing a high quality, variable centrality criterion.

\section{MULTIPLICITIES PER PARTICIPANT}\label{IV}

	 In order to compare the multiplicities measured with different beam-target combinations, 
we can use the calculated number of participants to determine the multiplicity per participant for 
each of our four beam-target systems.  Table III shows the average shower particle multiplicities 
per participant $m_s = n_s /W$ for the four central data sets of Table II. The first three systems 
are at 200 GeV/nucleon, and the Pb+Pb is at 158 GeV/nucleon. In the framework of the wounded 
nucleon model, the average number of produced particles per participant $m_{prod} = n_{prod} /W$ 
is independent of impact parameter.  To obtain the quantity $n_{prod}$,  the measured quantity 
$n_s$ is corrected by subtracting the number of spectator protons and adding the number of 
produced slow ($\beta < 0.7$) particles. The average number of projectile spectator protons 
$n_{spect}$ is estimated from the Glauber calculations.  The average number of heavily ionizing 
slow produced particles is estimated from FRITIOF or VENUS.

	Since the FRITIOF and VENUS predictions for the number of slow particles are different, 
we obtain two values for $m_{prod}$: one corrected with FRITIOF and one corrected with 
VENUS.  These are listed as $m_{prod}(Data)$  in Table III together with the two model 
predictions labeled $m_{prod}(Simulations)$. The two models give similar results for the heavy 
ion beams, but due to reinteraction, VENUS predicts a substantially larger multiplicity per 
participant for the proton beam than does FRITIOF.  Both model predictions slightly 
overestimate the respective measured multiplicities of the O+Ag/Br and S+Ag/Br, and predict 
significantly too many charged particles in Pb+Pb events.  FRITIOF's over-prediction of 
multiplicities in the Pb+Pb system is also discussed in Ref. [1].  However, we note that in the 
case of Pb + Ag/Br [27], where the target is significantly lighter than the projectile, the total 
measured multiplicities of high multiplicity events are consistent with the predictions.

	Table III also shows the charged multiplicities per participant $m_{pn} = n^{\pm}_{pn}/2$ for 
$pn$ interactions at similar energies [28,29].  We compare to $pn$ data rather than $pp$ data 
since the $pn$ 
system is more similar to the $AA$ data in isospin content and charge fraction, both of which 
affect the $NN$ multiplicities at the $10\%$ level.  FRITIOF and VENUS both produce almost 
equal numbers of protons and neutrons, which is not the case for $pp$ or $nn$ interactions, but 
is true of $pn$ interactions.  In agreement with other studies, we find that the data are 
systematically higher than predicted [5,6].  Table III appears to indicate that the Pb+Pb value is 
significantly closer to the wounded nucleon model prediction than are those for the other 
systems.  However, we note that the Pb+Pb system has a significantly different isospin mixture 
(i.e., a larger neutron excess) than the other systems, which may somewhat reduce its charged 
multiplicity relative to the other more proton-rich systems.  Nevertheless,  $m_{Pb+Pb}$ is lower 
than predicted by both Monte Carlos, and by empirical extrapolations from lighter systems 
[4,30].

Fig.~\ref{fig4} displays the shower multiplicity densities as a function of pseudorapidity.  For each 
system, the densities have been normalized by the calculated number of wounded nucleons for 
the data sample.  The same normalizations are used for the data and the models.  The error bars 
include uncertainties due to finite counting statistics and uncertainties in the systems' $b_{max}$ 
(which propagates into the uncertainty in $W$), added in quadrature.  The most obvious 
conclusion from this figure is that the Pb+Pb system has a significantly lower measured peak 
density than predicted by either model.  Qualitatively, this appears consistent with the trend from 
lighter to heavier systems.  The models slightly underestimate the p+Ag/Br central region, 
overestimate the O+Ag/Br and S+Ag/Br peaks, and significantly overestimate Pb+Pb.  FRITIOF 
provides the better fit in every case.  

	We have examined the effects of reinteraction predicted by VENUS with separate runs in 
which reinteraction has been turned off, without adjusting any other parameters in the model.  
These runs are represented in Fig.~\ref{fig4} by the dotted lines.  As expected, reinteraction improves the 
p+Ag/Br fit in the target region [31].  However, turning reinteraction off does not improve the fit 
to the AA data at midrapidities.  This result is discussed in more detail in the next section. 
 
	The hypothesis of limiting fragmentation [32,33] states that in hadronic interactions, the 
density in the target region is asymptotically independent of beam energy and projectile species.  
When the projectile region is measured in the target rest frame, it is shifted by a kinematically 
determined amount $y_{beam}$, but the projectile region densities are otherwise predicted to be 
independent of energy and target species.  This approximate invariance has been observed in 
$pp, \overline{p}p, pA$, and $\pi$-emulsion interactions [34-37].  The natural combination of 
this hypothesis with the principle of incoherent superposition in $AA$ collisions would state that 
the target region scales with the number of wounded target nucleons $W_T$ and the projectile 
region scales with the number of wounded projectile nucleons $W_P$ [38].  This expectation is 
tested in Fig.~\ref{fig5ab}, which shows the shower densities per wounded target nucleon in 
Fig. 5(a) and 
the densities per wounded projectile nucleon in Fig. 5(b).  Proton-hydrogen data from the NA22 
bubble chamber experiment [39] are shown for comparison to the emulsion data.  Fig.~\ref{fig5ab} 
demonstrates that wounded nucleon scaling is quite a good approximation only for $AA$ 
systems.  There appear to be some deviations from this scaling for $pp$ and $pA$ collisions, but 
these differences must be interpreted cautiously.  First, energy degradation of the projectile 
passing through the target is significantly different for $pp, pA$, and $AA$ interactions.  In 
addition, there are different spectator contributions.  The bubble chamber acceptance corrections 
are highest in the projectile region, and the acceptances for slow particles in the target region 
differ between the two techniques.  The p+Ag/Br data are lower than the O, S, and Pb data in the 
projectile fragmentation region, but the excesses in the $AA$ systems are consistent with the 
calculated number of spectator protons in these systems.  There are no projectile spectators in 
p+Ag/Br interactions.  The difference between the p+Ag/Br data and the $AA$ systems in the 
target regions appears to be real, and may be due to reinteraction.  Nevertheless, over a range of 
nearly two orders of magnitude in $W$ and $dn/d\eta$, the variation in $\frac{1}{W} dn/d\eta$ 
is only $\sim 0.3$. The pseudorapidity densities are parameterized quite well with just $W_T$ and 
$W_P$.  Dynamical effects apparently have only a minor effect on the angular distributions in 
AA systems in this energy range.

	In conclusion, multiplicities in the Pb+Pb system are lower than predicted either by 
FRITIOF or VENUS, although FRITIOF does significantly better than VENUS.  The 
discrepancy is primarily in the central region.  The distributions in the spectator regions appear to 
be reasonably well understood as the result of fragmentation of $W_P$ or $W_T$ wounded 
nucleons.  We observe no other dynamical effects in the spectator regions except in the p+Ag/Br 
system, where we may be observing the effects of reinteraction in the target region.

\section{SHAPES OF Pb+Pb PSEUDORAPIDITY DISTRIBUTIONS}\label{V}

	In the Pb+Pb system, the shapes of the shower pseudorapidity density distributions are 
independent of multiplicity over most of the observed region of pseudorapidity space. This behavior 
is illustrated in Fig.~\ref{fig6} using the full Pb+Pb data set.  The forward region $(\eta \geq 6)$ 
is the sole exception to this rule due to the presence of spectator protons, especially in the more 
peripheral (lower multiplicity) events.  Fig.~\ref{fig7} shows the total charge enclosed in the 
$\eta \geq 7$ and $\eta \geq 8$ cones.  (The $\eta \geq 6$ cone is shown in 
Fig.~\ref{fig3}.)  The relationship between 
enclosed charge and multiplicity is consistent with linearity in all three cones.  Thus, the data in 
the forward region are consistent with a linear two-component model, in which the produced 
multiplicity is directly proportional to multiplicity, just as in Fig.~\ref{fig6}, while the spectator 
contribution decreases linearly with multiplicity from the most peripheral events to the most 
central ones.  These linearities are predicted by FRITIOF, and are a consequence of 
superposition in symmetric systems.  They hold as long as second-order effects, i.e., 
reinteraction, are unimportant.

	VENUS models particle reinteraction, providing an expectation for the influence of reinteraction 
on the pseudorapidity distributions.  Fig.~\ref{fig8} relates the peak pseudorapidity density 
$(dn_s/d_{\eta})_{peak}$ of the Pb+Pb events to their total produced multiplicity $n_{prod}$. The 
straight line shows a linear fit to the data: The data are consistent with direct proportionality 
between $(dn_s/d_{\eta})_{peak}$ and $n_{prod}$.  The VENUS simulated events, superimposed on 
the data in Fig. 8(a), deviate from linearity and differ significantly in shape from the data at high 
multiplicity. This deviation can also be seen in Fig. 4, where VENUS predicts a narrower and 
taller distribution for central events than is observed.  In Fig. 8(b), the data points are suppressed 
for clarity, but the same straight line fit as in Fig. 8(a) is compared to VENUS with reinteraction 
turned off, but otherwise run with the same parameters.  VENUS is more consistent with the data 
when reinteraction is turned off, although at high multiplicities VENUS still predicts too high a 
central peak.  For the case of central Pb+Ag/Br collisions as well, a wider distribution than 
expected from VENUS was observed [27,40].  The best fit to the shape of the distribution is 
shown in Fig. 8(c), where the FRITIOF predictions are shown to be completely consistent with 
the linear fit to the data.

	The data are therefore consistent with two conclusions: (1) the shape of the produced 
particle multiplicity in the spectator region is independent of multiplicity, as it is from $\eta = 
2.9$ to $\eta = 6$, and (2) the shape of the spectator proton distribution is independent of 
multiplicity.  If these conclusions are correct, then we can statistically separate the produced and 
spectator distributions.  Let us define $f_{prod}(\eta)$ and $f_{spect}(\eta)$ to be the normalized 
produced particle and spectator proton distributions, respectively.  We normalize $f_{prod}(\eta)$ 
to integrate to one in the interval 2.9-5.5, a region which excludes essentially all spectators, and 
normalize $f_{spect}(\eta)$ such that it integrates to one over $\eta \geq 6$.  Then the average 
pseudorapidity distribution $\rho_{samp}(\eta)$ of a sample of events is

\begin{equation}
\rho_{samp}(\eta) = a_{samp}f_{prod}(\eta) + b_{samp}f_{spect}(\eta),
\end{equation}

\noindent where $b_{samp}$ is the number of spectator protons per event in the sample, and 
$a_{samp}$  is related to the produced multiplicity through the fraction of particles produced in the 
interval $\eta = 2.9-5.5$: i.e.,    $a_{samp} \sim 0.52 n_{prod}$.  If we measure the distributions of a 
central and a semi-central sample, $\rho_{cent}(\eta)$ and $\rho_{semi}(\eta)$, we can infer the 
shapes of the produced and spectator proton distributions.  For example,

\begin{equation} 
f_{spect}(\eta) = \frac{a_{semi}\rho_{cent}(\eta)-a_{cent}\rho_{semi}(\eta)} 
{a_{semi}b_{cent}-a_{cent}b_{semi}}
\end{equation}

\noindent Note that only $a_{semi}$ and $a_{cent}$ affect the shape of the resulting distribution; the 
numbers of spectators in the two samples $b_{semi}$ and $b_{cent}$ enter only into its 
normalization.  

	Our central sample for this analysis is the $Z_b \leq 4$  sample.  The semi-central sample 
consists of those events with $Z_b > 4$. The $a$ coefficients are simply the areas under the 
samples' measured distribution between $\eta = 2.9$ and $\eta = 5.5$.  The $b$ coefficients are 
evaluated using Fig. 3, which relates the mean multiplicity of each sample to its mean spectator 
charge $Q_{spect}$ in the $\eta \geq 6$ cone.  Subtracting the bound charge $Z_b$ gives us an 
estimate of the number of spectator protons in each sample, which is the same as $b_{samp}$.
  
	Fig. 9(a) shows the resulting spectator distribution $f_{spect}(\eta)$.  Fig. 9(b) shows the 
same distribution as a function of $\theta$  rather than $\eta$.  The $\theta$  distribution can be 
fitted to a gaussian with a half-width of $1.82\pm0.39$ mrad ($\chi^2 = 0.6$), as shown in Fig. 
9(b).  For spectators having the same longitudinal momentum as the beam, this width 
corresponds to an r.m.s. transverse momentum of $(290\pm60)$ MeV/c.\footnote{Presumably, the 
longitudinal momentum ($p_l$) distribution is at least as broad as the $p_t$ distribution (in the 
beam rest frame).  In this case, the lab-frame $p_l$ distribution will be quite broad, having a 
significant tail below 100 GeV/c.  Thus, our assumption that the spectator protons are a 
mono-energetic beam is not entirely realistic.  This may cause us to overestimate the r.m.s. transverse
 momentum.  In this case, the observed effect is due to a decreased longitudinal momentum of the 
spectators.}   This is significantly 
larger than would be expected from an isotropic evaporation model (140 MeV/c).  We can also 
fit the observed distribution to a sum of a Gaussian with a width of 140 MeV/c plus a second 
Gaussian peaked at $\theta \sim 3$ mrad, corresponding to $p_t \sim 470$ MeV/c.  In either case, 
there is evidence for a wider distribution than expected. Thus, the spectator distribution suggests 
that there is a rescattered component.  This scattered component, however, is small: the 
contribution of a second Gaussian to the total is no more than $\sim 20\%$ of the total.  

	The derived spectator distribution is consistent with our other results.  The distribution has 
a small tail extending beyond $\eta = 6$ ($\theta \sim 5$ mrad), indicating that the spectator 
charge residing outside the $\eta \ge 6$ cone is very small, in agreement with the charge intercept in Fig. 
3.  The derived produced particle distribution has almost the same shape as that of the central 
sample, but has a slightly smaller tail, as expected (Fig.~\ref{fig10}).  We note that the spectator 
correction improves the agreement between the Pb+Pb and the p+Ag/Br projectile regions in Fig. 
5.

	In summary, the data are consistent with the shape of $dn_{prod}/d\eta$ being independent 
of multiplicity over the entire range of pseudorapidity.  This independence is directly observed in 
the central region, where there is no spectator contribution.  This result is unexpected in light of 
the VENUS simulations, and appears to indicate that reinteraction has little influence on the 
produced particle distribution.  On the other hand, the typical $p_t$ values derived for spectator 
protons are between typical Fermi momenta ($\sim 100$ MeV/c) and proton scattering $p_t$ values 
($\sim 450$ MeV/c), suggesting that reinteraction plays a role in scattering or heating the residual 
spectators.

\section{DISCUSSION AND CONCLUSIONS}\label{VI}

	Among high energy heavy ion systems studied to date, the 158 GeV/nucleon Pb+Pb system 
is unique in its combination of symmetry with large multiplicities, and consequently high track 
statistics in individual events.  We have exploited both of these properties in studying centrality 
criteria, forward multiplicities, and shape-multiplicity dependence. Because of the symmetry of 
the Pb+Pb system, any shape changes in the pseudorapidity distributions must be due to 
reinteraction or the onset of non-superposition effects, rather than changes in collision geometry.  
However, we find no evidence for shape changes.  Our analyses by no means rule out the 
occurrence of reinteraction, but taken together, they place stringent limits on  rescattering effects 
in the Pb+Pb system as they might appear in multiplicity or pseudorapidity density 
measurements.  Our study of forward multiplicities of individual events gives us confidence in 
our participant estimates, and also allows us to plausibly extract information on the transverse 
momenta of spectator protons.  These features of the dataset are summarized in Fig. 3, which 
brings together the relationships between the multiplicities of central events, forward charge, 
number of participants, and beam charge.

	Perhaps the most interesting finding presented here is the independence of the 
pseudorapidity shape and the centrality of the collision. This apparently implies that at SPS 
energies, particle production in the center of mass is not significantly more isotropic in central 
events than it is in peripheral ones, even in ultra-heavy systems.  This contrasts with the results 
from the AGS at 14.6 GeV/nucleon, where pseudorapidity densities from heavy ion interactions 
become roughly isotropic at the highest multiplicities [41]. Near 200 GeV/nucleon, this 
anisotropy holds not only in the central region but also in the spectator regions, where we 
observe scaling with the number of participating projectile or target nucleons: the initial 
geometry of the system is reflected in the final state.  

	The fact that rescattering-induced narrowing is predicted to occur but is not observed is 
puzzling in light of evidence for rescattering from other experiments [42-44].  In VENUS, the 
narrowing of the distribution with reinteraction occurs because of a combination of greater 
proton stopping power, slower pions, and enhanced heavy particle production ($\overline{p}p$ pairs and 
kaons) in the central region.  Fig.~\ref{fig11} shows the interplay of these effects.  Note that the 
energy for additional heavy particle production comes in part from increased nucleon slowing, 
and also a slight reduction in pion production.  The absence of narrowing in central events may 
indicate that the degree of slowing of protons and pions in Pb+Pb events is less than the model 
predicts.  Without the additional energy available for particle production which comes from 
increased nucleon stopping power, increased heavy particle production would come at the 
expense of smaller pion yields.  Thus, reinteraction-induced heavy particle production 
accompanied by a more modest increase in stopping power could explain both the observed 
shape independence and the low multiplicities.

	In conclusion, it seems clear that in general FRITIOF provides a better simulation of the 
data than does VENUS. This is particularly noticeable for these very massive Pb+Pb 
interactions, and becomes less noticeable for the lighter nuclei.  The average multiplicities and 
pseudorapidity distributions in central AA interactions on heavy targets at SPS energies are 
mainly determined by the number of participating nucleons. Multiplicities in these collisions are 
nonetheless higher than would be expected by simply scaling NN multiplicities at the same 
energy, indicating a dynamical effect at work in addition to the wounded nucleon effect.  The 
dependence of this excess on system mass is not well-predicted by the Monte Carlo event 
generators.  The Pb+Pb pseudorapidity distributions appear to have a shape which is independent 
of multiplicity, placing an important constraint on models of reinteraction in the central region.  
Indeed, no direct evidence for reinteraction is observed in the central region.  However, we do 
find evidence for reinteraction of produced particles or excited matter on the spectators.

\acknowledgments 
	This work was partially funded in the U.S. by the National Science Foundation (Grants 
Nos. PHY-9513997 and INT-8913051 at LSU) and Department of Energy (Grant No. DOE-
FG02-89ER40528 at Minnesota), and in Poland by State Committee for Scientific Research 
Grant No. 2P03B05417 and Maria Sklodowska-Curie Fund II no. PAA/NSF-96-256.  P.D. 
thanks the Louisiana State Board of Regents (LEQSF) under agreements Nos. NASA/LSU-91-
96-01 and NASA/LaSPACE under Grant No. NGT-40039 for its support.  We appreciate the 
help of the CERN staff, A. Aranas, J. Dugas, and L. Wolf at LSU, and especially thank Professor 
Y. Takahashi and his EMU-16 colleagues for their generous assistance.

\newpage

\begin{figure}
\caption{Correlation between the darkness measurements on the Pb projectile fragments and their 
charges estimated from the number of $\delta$-rays counted along the fragment tracks in the 
charge section of the EMU-13 chambers.}\label{fig1}
\end{figure}

\begin{figure}
\caption{a) Measured mean multiplicity and b) standard deviation of Pb+Pb events selected by 
the charge $Z_b$ bound in the multiply charged projectile fragments(solid circles).  Simulated 
events are selected by impact parameter (open symbols).}\label{fig2}
\end{figure}

\begin{figure}
\caption{Forward charge vs. total multiplicity of produced particles (open circles).  Large solid 
circles represent the averages for samples with $n_{prod} \geq 1000$ and $Q_{\eta \geq 6} = 0, 2, 3-
4, 5-7, and \geq 8$.  Large solid triangles are the calculated produced forward multiplicity 
$n_{\eta>6} = Q_{\eta \geq 6} - Q_{spect}$, where $Q_{spect}$ is the spectator charge derived from a 
Glauber calculation.  The upper line is the fit to the data.  The intercept is consistent with 
$Z=82$, the charge of the beam.  The lower line is fitted to the forward multiplicity points and is 
constrained to pass through the origin.  The intersection of the two lines is an estimate of the 
multiplicity in events with no spectators.}\label{fig3}
\end{figure}

\begin{figure}
\caption{Densities of shower particles per wounded nucleon $dm_s/d\eta$ as a function of the 
pseudorapidity $\eta$.  The error bars include statistical counting uncertainties in $W$.  For each 
particular system, the same values of $W$ are used to normalize the data, FRITIOF(dashed 
lines), and VENUS. VENUS predictions are shown with (solid lines) and without (dotted lines) 
reinteraction.}\label{fig4}
\end{figure}

\begin{figure}
\caption{Scaling of target and projectile regions with the number of target and projectile 
participants.  Note that the two vertical axes are different.  The target region (a) is normalized by 
$W_T$, while the projectile region (b) is normalized by $W_P$.}\label{fig5ab}
\end{figure}

\begin{figure}
\caption{Dependence of pseudorapidity density with multiplicity in three pseudorapidity 
intervals.  The fitted lines are constrained to pass through the origin.}\label{fig6}
\end{figure}

\begin{figure}
\caption{Comparison of the charge forward of (a) $\eta=7$ and (b) $\eta=8$ as a function of the 
total produced multiplicity to a linear model.}\label{fig7}
\end{figure}

\begin{figure}
\caption{Comparison of measured peak pseudorapidity densities $(2.9 \leq \eta <3.6)$ with 
VENUS predictions (a) with reinteraction included, and (b) with reinteraction turned off but 
with all other parameters left unchanged.}\label{fig8}
\end{figure}

\begin{figure}
\caption{Derived spectator proton distribution in the Pb+Pb system. The error bars shown are 
statistical.  Fig. 9(a) shows the distribution as a function of pseudorapidity $\eta$ derived from 
Eq. 4;  Fig. 9(b) shows the distribution transformed into angular units.  In Fig. 9(b), the $\eta>9$ 
data, which has large relative uncertainties in the angle, is not plotted or used in the fit.}
\label{fig9}
\end{figure}

\begin{figure}
\caption{Comparison of the derived produced particle distribution as a function of 
pseudorapidity $\eta$ with the central ($Z_b \leq 4)$ sample in the spectator region.}\label{fig10}
\end{figure}

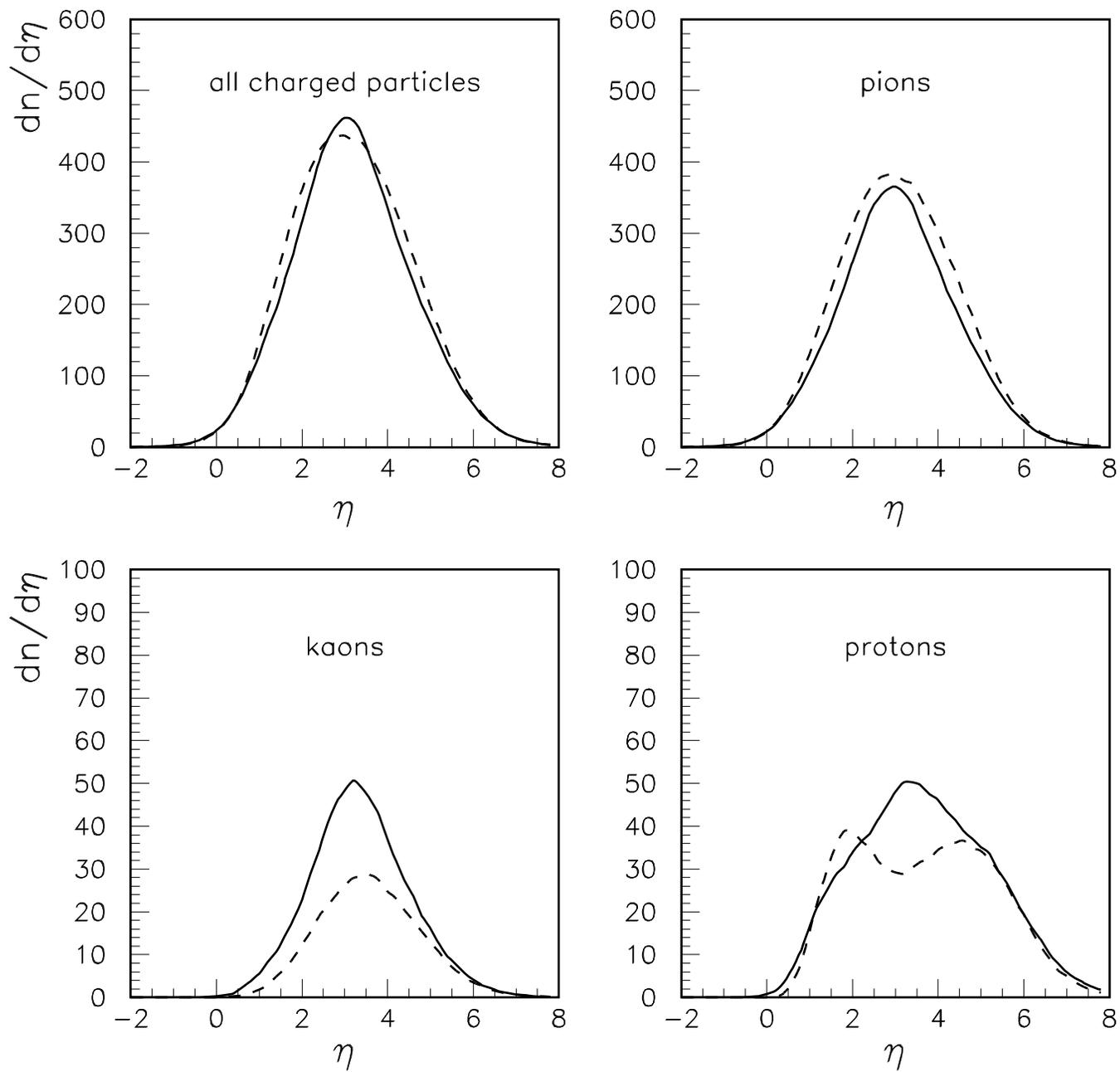
\begin{figure}
\caption{Pseudorapidity distributions of dominant charged particle species in VENUS, with 
(solid lines) and without (dashed lines) reinteraction.}\label{fig11}
\end{figure}

\newpage

\begin{table}
\caption{158 GeV/nucleon  Pb+Pb samples}\label{TableI}
\begin{tabular}{lccccc} \tableline
Sample & Events & $<Z_b>$ & $<n_{prod}>$ & $<n_{fr}>$ & $<Q_{\eta\ge6}>$ \\ \hline
$Z_b = 0$ & $21$ & $0$ & $1400\pm57$ & $0$ & $50.7\pm1.8$\\
$Z_b = 2$ & $25$ & $2$ & $1241\pm50$ & $1$ & $53.0\pm1.8$\\
$Z_b = 3 - 6$ & $24$ & $4.8\pm0.2$ & $1096\pm48$ & $2.33\pm0.10$ & $57.0\pm1.8$\\
$Z_b = 7 - 12$ & $32$ & $9.4\pm0.3$ & $899\pm39$ & $3.94\pm0.20$ & $62.8\pm1.3$\\
$Z_b = 13 - 17$ & $23$ & $14.5\pm0.3$ & $779\pm36$ & $5.96\pm0.19$ & $67.3\pm1.6$\\
$Z_b = 18 - 25$ & $24$ & $21.0\pm0.5$ & $651\pm36$ & $7.88\pm0.31$ & $69.0\pm1.8$\\
$Z_b > 25$      & $21$ & $33.2\pm1.2$ & $468\pm29$ & $8.67\pm0.63$ & $71.1\pm1.6$\\ \hline
$n_{prod} \ge 1000$ & $71$ & $3.2\pm0.4$ & $1263\pm43$ & $1.46\pm0.17$ & $53.3\pm1.0$\\ \hline 
total & $170$ & $11.8\pm0.8$ & $933\pm38$ & $4.21\pm0.25$ & $61.5\pm0.8$\\
\end{tabular}
\end{table}
			
\begin{table}
\caption{Proton-nucleus and nucleus-nucleus central interactions used in the analysis}
\label{TableII}
\begin{tabular}{lccccccc} \tableline
System & Event & $N_{cent}/N_{tot}$ & $<n_s>$\footnote{Errors are statistical, except for Pb+Pb, 
       which includes systematics.} & $b_{max}$(fm) & $W_P$ & $W_T$ & $\nu$\footnote{The total number 
       of interactions $\nu = W_T\nu_P =W_P\nu_T$ , where for example 
       $\nu_T = A_T\sigma_{nn}/\sigma_{nT}$.} \\ \hline
p(200)+Ag/Br & $451$ & $24.4\%$ & $22.4\pm0.4$ & $2.89\pm0.13$ & $1$ & $4.2\pm0.1$ & $4.2$\\
O(200)+Ag/Br & $151$ & $28.4\%$ & $172\pm4$    & $4.3\pm0.4$   & $14.6\pm0.4$ & $27.5\pm1.4$ & $50$\\
S(200)+Ag/Br & $472$ & $19.8\%$ & $288\pm4$    & $4.0\pm0.4$   & $28.4\pm0.8$ & $42.2\pm2.1$ & $97$\\
Pb(158)+Pb   & $71$  & $9.7\%$ & $1263\pm43$\footnote{Produced multiplicity $n_{prod}$.  See text.}
                                          & $4.6\pm0.6$  & $170\pm7$    & $170\pm7$   & $767$\\ 
\end{tabular}
\end{table}

\begin{table}
\caption{Charged multiplicities}\label{TableIII}
\begin{tabular}{lccccccc} \tableline
 & & & \multicolumn{2}{c}{$m_{prod}(Data)$} & \multicolumn{2}{c}{$m_{prod}(Simulations)$} & \\ 		
System  & $m_s$ & $n_{spect}$ & VENUS & FRITIOF & VENUS & FRITIOF & $m_{pn}$ \\
p+Ag/Br & $4.3\pm0.1$ & $0$ & $5.5\pm0.1$ & $4.5\pm0.1$ & $5.3\pm0.1$ & $3.9\pm0.1$ & $3.74\pm0.03$ \\
O+Ag/Br & $4.1\pm0.2$ & $1.0$ & $4.4\pm0.2$ & $4.2\pm0.2$ & $4.8\pm0.1$ & $4.6\pm0.1$ & $3.74\pm0.03$ \\
S+Ag/Br & $4.1\pm0.1$ & $2.2$ & $4.3\pm0.1$ & $4.1\pm0.1$ & $4.8\pm0.1$ & $4.7\pm0.1$ & $3.74\pm0.03$ \\
Pb+Pb	 & $3.7\pm0.2$ & $16.0$ & $3.7\pm0.2$ & $3.7\pm0.2$ & $4.7\pm0.1$ & $4.3\pm0.1$ & $3.48\pm0.04$ \\
\end{tabular}
\end{table}

\end{document}